\documentclass[10pt, conference]{IEEEtran}
\usepackage[utf8]{inputenc}
\usepackage[detect-all]{siunitx}
\usepackage{array,tabularx,calc}
\usepackage{mathtools}
\usepackage{amsfonts}
\usepackage{graphicx}
\usepackage{amssymb}
\usepackage{amsfonts}
\usepackage{pbox}
\usepackage{url}
\usepackage[]{algorithm, algpseudocode}
\usepackage{tikz}
\usepackage{url}

\DeclareSIUnit{\belmilliwatt}{Bm}
\DeclareSIUnit{\dBm}{\deci\belmilliwatt}
\DeclareSIUnit{\belisotropic}{Bi}
\DeclareSIUnit{\dBm}{\deci\belisotropic}
\DeclareSIUnit{\bit}{bit}

\ifCLASSOPTIONcompsoc	\usepackage[caption=false,font=normalsize,labelfont=sf,textfont=sf]{subfig}
\else					\usepackage[caption=false,font=footnotesize]{subfig}
\fi

\makeatletter
\usepackage[]{algorithm, algpseudocode}
\makeatletter
\renewcommand{\ALG@beginalgorithmic}{\small}

\usepackage[capitalise]{cleveref}	
    \crefformat{equation}{(#2#1#3)}	

\usepackage[noadjust]{cite}

\begin{document}

\usetikzlibrary{arrows}
\usetikzlibrary{shapes}
\newcommand{\mymk}[1]{%
	\tikz[baseline=(char.base)]\node[anchor=south west, draw,rectangle, rounded corners, inner sep=0.1pt, minimum size=3.5mm,
	text height=2mm](char){\ensuremath{#1}} ;}

\newcommand*\circled[1]{\tikz[baseline=(char.base)]{
	\node[shape=circle,draw,inner sep=0.1pt] (char) {#1};}}

\title{ A Framework to Develop and Validate RL-Based Obstacle-Aware UAV Positioning Algorithms}

\author{\IEEEauthorblockN{Kamran Shafafi, Manuel Ricardo, Rui Campos }
	\IEEEauthorblockA{INESC TEC and Faculdade de Engenharia, Universidade do Porto, Porto, Portugal\\
		\{kamran.shafafi, manuel.ricardo, rui.l.campos\}@inesctec.pt}}

\maketitle

\begin{abstract}

Unmanned Aerial Vehicles (UAVs) increasingly enhance the Quality of Service (QoS) in wireless networks due to their flexibility and cost-effectiveness. However, optimizing UAV placement in dynamic, obstacle-prone environments remains a significant research challenge due to their complexity. Reinforcement Learning (RL) offers adaptability and robustness in such environments, proving effective for UAV optimization.

This paper introduces RLpos-3, a novel framework that integrates standard RL techniques and simulation libraries with Network Simulator 3 (ns-3) to facilitate the development and evaluation of UAV positioning algorithms. RLpos-3 serves as a supplementary tool for researchers, enabling the implementation, analysis, and benchmarking of UAV positioning strategies across diverse environmental conditions while meeting user traffic demands. To validate its effectiveness, we present use cases demonstrating RLpos-3's performance in optimizing UAV placement under realistic conditions, such as urban and obstacle-rich environments.

\end{abstract}


\section{Introduction}

Unmanned Aerial Vehicles (UAVs) offer unique capabilities, including mobility, cost-effectiveness, and flexible positioning anywhere and at any time. Compared to terrestrial infrastructure, UAV-based networks provide greater agility and ease of configuration. This is particularly beneficial in disaster management scenarios, such as wildfires, earthquakes, floods, cyberattacks, and terrorist attacks, where conventional infrastructure may fail \cite{shafafi2023uav, shafafi2024traffic}. Moreover, during festivals and crowded events, UAVs can enhance network capacity or temporarily replace traditional infrastructure as needed \cite{shafafi2023joint, jx7c-py87-25}. In commercial and civilian domains, UAVs enable applications such as weather monitoring, forest fire detection, traffic control, cargo transport, emergency search and rescue, and communications relay. Consequently, UAVs have gained significant traction in recent years for deploying mobile Base Stations (BSs) and Wi-Fi Access Points (APs) \cite{rs11121443}. These capabilities make UAVs critical for enhancing wireless network coverage in today’s digital society, supporting services like augmented reality, online gaming, ultra-high-definition video streaming, disaster safety, and event facilitation \cite{8875210}.

UAV deployment faces challenges, including environmental conditions, distance, attenuation, obstacle effects, user traffic demands, transmission power, and energy efficiency. Properly deploying UAVs at precise coordinates in real-world scenarios requires a flexible positioning system adaptable to diverse conditions. Although various studies have investigated UAV positioning, most solutions propose algorithms, such as optimization techniques, heuristic methods, and, recently, Machine Learning (ML) approaches, tailored to specific cases. In complex, obstacle-rich scenarios, relying on optimization and heuristic methods may be insufficient. Consequently, a fundamental gap remains due to the absence of a generic framework capable of addressing challenges across multiple environments. Reinforcement Learning (RL) has emerged as an effective approach, providing adaptability and robustness in dynamic, obstacle-rich contexts. Although prior work optimizes UAV positioning in obstacle-rich urban environments using a customized Deep Q-Network (DQN) algorithm \cite{Reinforc82:online}, it lacks flexibility for other contexts. To address this gap, RLpos-3 provides a modular framework adaptable to both obstacle-rich and free-space scenarios, supporting user-configurable environments, algorithms, and performance metrics.

\begin{figure}
	\centering
		\includegraphics[width=\linewidth]{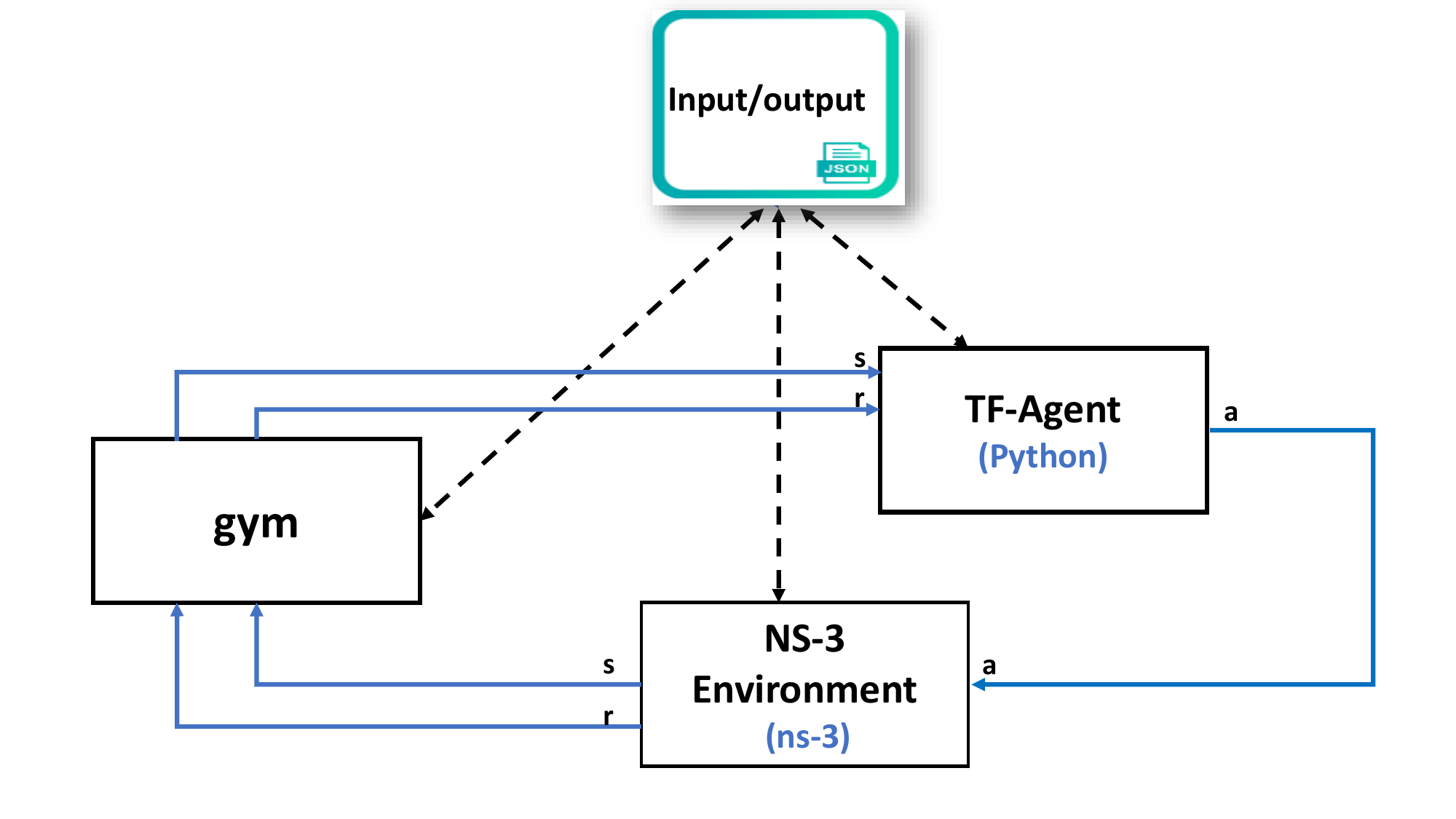}
	\caption{RLpos-3 main modules and interactions between them.}
	\label{fig1}
\end{figure}

The main contribution of this paper is RLpos-3, a generic and modular simulation framework designed to implement, validate, and evaluate RL-based UAV positioning algorithms in aerial networks. RLpos-3 integrates standard RL libraries, such as TensorFlow~Agents \cite{TensorFl16:online} and OpenAI~Gym \cite{GitHubop15:online}, with Network~Simulator~3 (ns-3.44) \cite{ns3adisc37:online}. It customizes these libraries to meet UAV positioning requirements in diverse scenarios, using the ns3-gym interface \cite{GitHubtk28:online} as a communication bridge between RL agents and the ns-3 simulator. By leveraging RL, RLpos-3 achieves enhanced positioning accuracy, reduced latency, and improved network throughput through line-of-sight (LoS) links with ground users. Unlike algorithm-specific implementations tailored to fixed settings, RLpos-3 is scenario-agnostic, supporting both obstacle-rich and open-space environments and accommodating arbitrary numbers of ground users and buildings as obstacles. It allows the integration of various RL algorithms (e.g., DQN, Proximal Policy Optimization (PPO), and Soft Actor-Critic (SAC)) and enables custom reward functions. A key feature of RLpos-3 is its traffic-awareness, dynamically optimizing UAV positions using RL to meet user traffic demands and evaluating the resulting placements. The source code of RLpos-3 is publicly available to the research community\footnote{``RLpos-3: Reinforcement Learning for Positioning.'' IEEE DataPort, \url{https://dx.doi.org/10.21227/rp8v-jd54}} \cite{RLpos3Re43:online}.

The rest of this paper is organized as follows. Section \ref{sec:soa} provides an overview of state-of-the-art positioning approaches in flying networks. Section \ref{sec:RLpos-3 Framework} provides a detailed description and structure of RLpos-3. Section \ref{sec:performance_evaluation} presents the performance evaluation of RLpos-3, including a sample simulation setup, performance metrics, and results. Section \ref{sec:Conclusions} summarizes the main conclusions and suggests directions for future research. research.

\section{State of the Art~\label{sec:soa}}

Many studies address the UAV positioning problem through optimization-based and mathematical formulations. However, these works often overlook real-world deployment challenges and are limited to specific, idealized scenarios. A significant number of studies assume simplified environment -- such as obstacle-free area -- which limit their applicability in practical deployments. This section reviews representative studies on UAV positioning algorithms and their tools, emphasizing the need for generalizable solutions.

Optimization-based approaches dominate UAV positioning research, often assuming simplified environments. In \cite{8761897}, the authors propose an optimization framework to enhance uplink throughput and resource allocation, assuming an obstacle-free environment. Similarly, \cite{8422376} focuses on optimizing the backhaul network while maximizing ground user coverage. The authors of \cite{9684492} present a UAV-aided ground positioning method using a nonparametric Belief Propagation (NBP)-based probabilistic framework to improve UAV localization while minimizing interference. The study in \cite{8369021} optimizes the number of UAVs required to cover a designated area, also under simplified conditions. For indoor scenarios, \cite{7921981} proposes a solution in which a single UAV provides wireless connectivity to UEs within a building. Meanwhile, \cite{7510820} investigates 3D UAV placement to maximize network revenue, using a bisection search algorithm that determines the UAV’s coverage area and altitude simultaneously.

More recently, ML and RL approaches have emerged as promising alternatives for UAV positioning. For example, \cite{ML8121867} and \cite{MLJiang2017MachineLP} explore ML-based positioning strategies tailored to specific scenarios, yet lack general-purpose frameworks. In \cite{43-bab6dcd317da4e3d82375669cbb45023}, deep learning maximizes network throughput, while \cite{23-8644345} proposes a three-step method for 3D UAV positioning and mobility in multi-UAV systems. The work in \cite{22-COLONNESE2019101872} employs Q-learning to enhance Quality of Service (QoS) and Quality of Experience (QoE) in UAV networks. These studies underscore the need for a unified RL-based framework like RLpos-3.

In summary, although existing solutions offer valuable insights for targeted scenarios, a generalizable and modular framework for UAV positioning remains largely unexplored. Optimization-based approaches often lack scalability and adaptability across diverse conditions. In contrast, Artificial Intelligence (AI), particularly RL, provides a more flexible and robust alternative. To the best of our knowledge, RLpos-3 is the first unified simulation framework integrating Network Simulator 3 (ns-3.44) with OpenAI~Gym and TensorFlow~Agents for UAV positioning. Its Environment module models obstacle-rich environments -- via the $BuildingModule$ -- and free-space scenarios, with configurable parameters such as venue size ($S_{\text{venue}}$) and obstacle density. This modular design allows for flexible testing, validation, and benchmarking of RL-based UAV positioning strategies under a wide range of network conditions.

\section{RLpos-3 Framework~\label{sec:RLpos-3 Framework}}

The RLpos-3 framework, designed for UAV positioning, comprises four main modules: $Input/output$, $Environment$, $Agent$, and $gym$. Figure \ref{fig1} illustrates these modules and their interactions. The \emph{Input/output} module, a JSON-based script, interfaces with users, manages configuration parameters, and enables communication with other modules. The \emph{Environment} module defines the state (\( s \)) of the RL process. At each timeslot \( t_s \), the \emph{Agent}, via the \emph{gym} interface, receives state \( s \) from \emph{Environment}, executes action \( a \), and receives a reward \( r \), relayed by \emph{gym} as feedback. Based on \( r \), the \emph{Agent} selects the next action for the subsequent \( t_s \). The \emph{Input/output} module also defines essential RL hyperparameters, including training episodes, evaluation episodes, batch size, learning rate, epsilon-greedy value, and buffer size.

\begin{figure}
	\centering
		\includegraphics[width=0.9\linewidth]{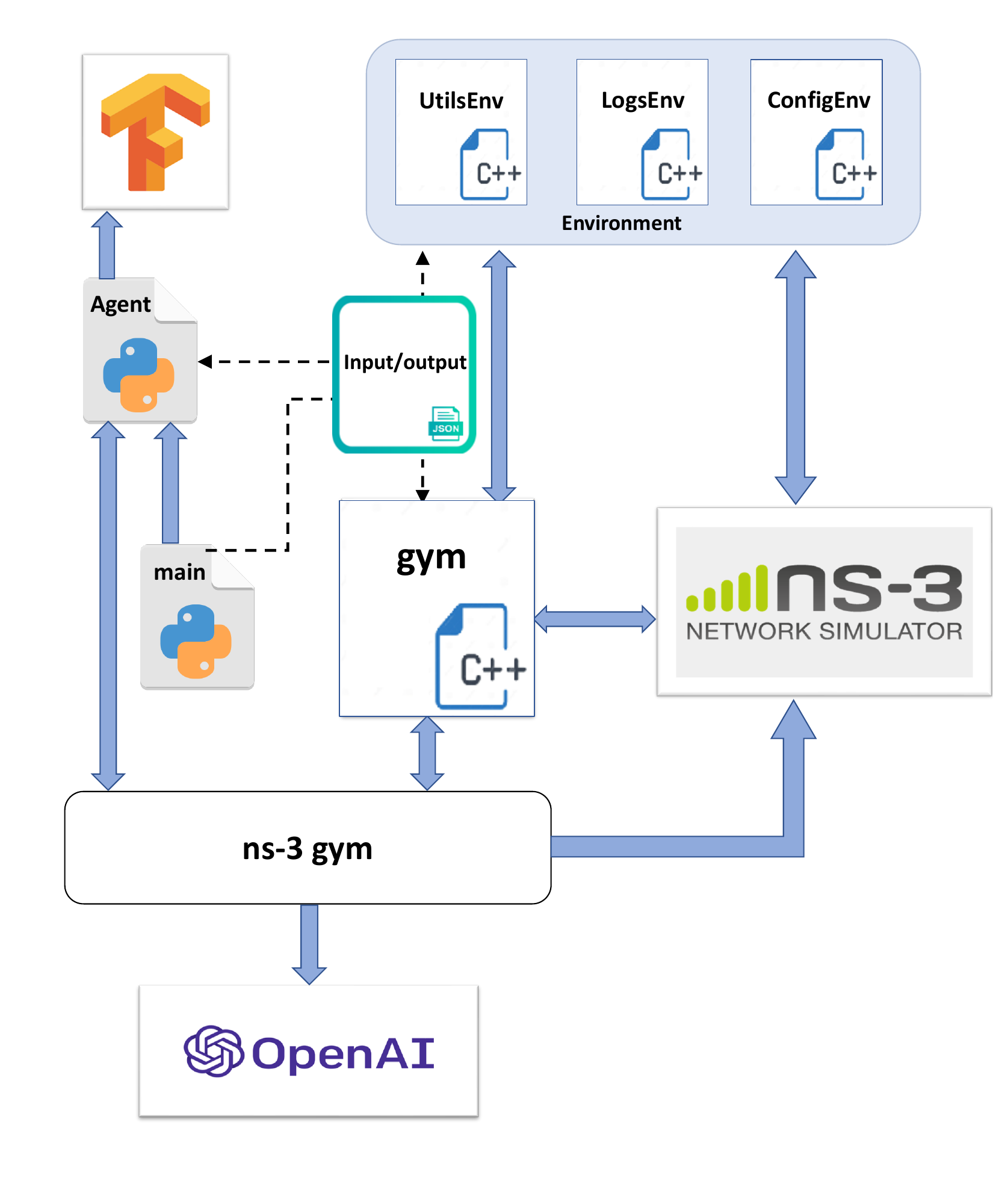}
	\caption{RLpos-3 architecture block diagram.}
	\label{fig2}
\end{figure}

Figure \ref{fig2} illustrates the architecture and submodules developed within each module. While ns-3 provides the core network simulation capabilities and protocols the \emph{Environment} module, developed using ns-3, extends its capabilities to simulate wireless network environments with UAVs and ground users, including wireless configurations, mobility patterns, and obstacle-rich environments. It includes three submodules: $UtilsEnv$ manages utility functions for configuration, Wi-Fi channel setup, and command-line interface operations; $ConfigEnv$, the core configuration module, handles network parameters (e.g., Wi-Fi settings, mobility, application deployment), supports User Datagram Protocol (UDP) and Transmission Control Protocol (TCP) selection via the \emph{Input/output} module, and manages building implementation and node positioning; $LogsEnv$ provides comprehensive logging across components, with configurable debug levels (Error, Warning, Info, Debug) based on \emph{Input/output} settings, facilitating error detection and correction.

Due to ns-3's modeling constraints, RLpos-3 considers only buildings as obstacles. The system supports buildings with varying heights, floors, and room sizes. Users of RLpos-3 define building coordinates and venue dimensions via the \emph{Input/output} module. RLpos-3 supports Wi-Fi communication, allowing selection among IEEE 802.11 standards and providing support for various remote station manager mechanisms. The Wi-Fi medium operates in either STA-AP or Ad Hoc mode, configured through the \emph{Input/output} module. Propagation loss models are selected based on environmental conditions. Users select loss models via the \emph{Input/output} module, and the RLPOS-3 framework will adjust accordingly. The \(FriisPropagationLossModel\) is suitable for obstacle-free scenarios. In environments with obstacles and Non-Line-of-Sight (NLoS) connections, models like \(HybridBuildingsPropagationLossModel\), \(ItuR1411LosPropagationLossModel\), and \(ItuR1411Nlos-OverRooftopPropagationLossModel\) are available.

The \emph{gym} module, built on ns3-gym, extends it to create a specialized environment for UAV positioning optimization. While ns3-gym bridges OpenAI Gym and ns-3, \emph{gym} customizes the interface for UAV-based wireless network optimization. The main customizations are:

\begin{itemize}
     \itemsep0em
    \item \textbf{MyGymEnv:} A customized \emph{gym} environment that inherits from OpenGymEnv, the standard \emph{gym} interface. This environment extends OpenGymEnv by implementing methods and parameters for UAV positioning
    \item \textbf{ScheduleNextStateRead:} Schedules the timestep \( t_s \). \( t_s \) is the time considered by the agent to execute new actions and collect state
    \item \textbf{GetActionSpace:} An OpenAI Gym discrete space that defines movement directions for the agent, configured via the \emph{Input/output} module, including default UAV directions (up, down, forward, backward, left, right, stay). The dimensions and types of actions are flexible and can be configured based on the target scenario
    \item \textbf{GetObservationSpace:} Configures an OpenAI Gym box space for observable state parameters, defined via the \emph{Input/output} module, including default parameters (UAV position, throughput, LoS connections). The configuration enables any type and number of observations
    \item \textbf{ExecuteActions:} The agent executes actions during each time step, with adjustable positioning zone limits aligned with venue scale or building heights in obstacle-rich scenarios
    \item \textbf{ReceivePacketRX:} Monitors the reception of packets
    \item \textbf{ThroughputMonitor:} Computes the throughput
    \item \textbf{DelayMonitor:} Observes the mean delay of individual packets
    \item \textbf{TrafficDemandMonitor:} Assesses whether the potential UAV position proposed by the agent can meet the traffic demands of all UEs. The mechanism applies a Modulation Code Scheme (MCS) for each UE based on demanded traffic to enforce a minimum Signal-to-Noise Ratio (SNR)
    \item \textbf{GetGameOver:} Checks for completion of training and evaluation
\end{itemize}

The \emph{Agent} module, developed using the TensorFlow Agents (TF-Agents) framework, extends its Deep Q-Network (DQN) implementation to interact with the \emph{gym} module (built on ns3-gym). While TF-Agents provides the core RL algorithms and neural network architectures, this module includes customizations for UAV positioning, such as tailored training and evaluation procedures. The framework is executed via a Python script named \(main.py\). Users can select training or evaluation mode. Main \emph{Input/output} module parameters are specified via the command-line interface.

\section{Applying RLpos-3: Example-Based Testing in Action~\label{sec:performance_evaluation}}

This section presents a usage example of RLpos-3 for training and evaluating a UAV positioning algorithm across diverse scenarios. The objective is to optimally position the UAV to ensure LoS connectivity with all UEs while maximizing throughput and accommodating each UE's traffic demand. RLpos-3 is used to train and validate the algorithm, aiming to enable reliable broadband communication at higher frequencies. The achieved solution is then evaluated in terms of QoS, particularly throughput and delay, providing valuable insights into the performance, functional validation, and adaptability of RLpos-3 under different environmental and traffic demand conditions. The findings are also generalizable to more complex scenarios involving varying numbers of users and obstacles.

\begin{figure}
	\centering
		\includegraphics[width=0.8\linewidth]{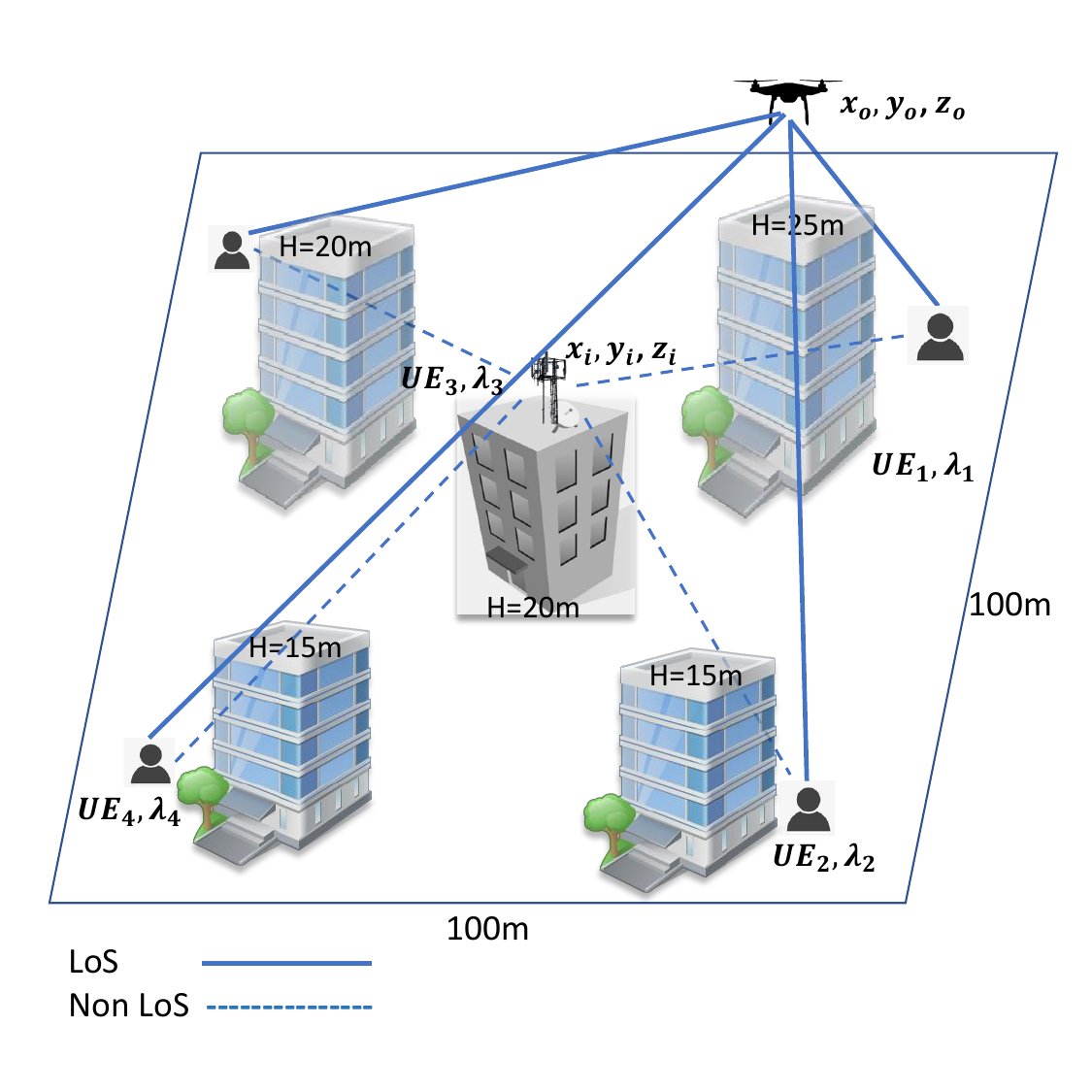}
	\caption{Evaluation scenario, which involves four UEs positioned in non-LoS locations with a UAV. ($x_o, y_o, z_o$) is the optimal position, achieved by RLpos-3 for the UAV to establish LoS connections with all four UEs.}
   \label{fig3}
\end{figure}

\subsection{System Settings~\label{System Settings}}

To evaluate the performance of RLpos-3, we considered three different scenarios: A) a free space scenario where 20 UEs are distributed within a 100m x 100m venue with constant mobility and each demanding a traffic rate of $\lambda = 58.5$ Mbit/s associated with the MCS index 0; B) an obstacle-rich homogeneous scenario includes five buildings of varying heights and four UEs with constant mobility in the same size venue where $\lambda_0 =  \lambda_1 =  \lambda_2 =   \lambda_3$, associated with the MCS index 0, with traffic demand of 58.5 Mbit/s; C) an obstacle-rich heterogeneous scenario includes five buildings of varying heights and four UEs with constant mobility in the same size venue where  $\lambda_0 = 0.75 \times \lambda_1 = 2 \times \lambda_2 = 4 \times \lambda_3$, with traffic demand of 234, 175.5, 117, and 58.5 Mbit/s associated with the MCS index 3, 2, 1, and 0, respectively, as depicted in Figure \ref{fig3}. Consider that the UEs are placed at the coordinates \((x_u, y_u, z_u)\), where \(u \in \{0,...,3\}\). The UAV is equipped with a Network Interface Card (NIC) operating in ad hoc mode, utilizing the IEEE 802.11ac standard on channel 50, with a channel bandwidth of 160 MHz and a Guard Interval (GI) of 800 ns. A single spatial stream is used for all links between the UEs and the UAV. Each UE generates UDP traffic using the $OnOffApplicationModule$, directed to the UAV, which has a UDP sink receiver installed. Each UE is assigned a traffic demand \(\lambda_u\), where \(u \in \{0,...,3\}\), corresponding to a specific MCS index \cite{MCSTable16:online}. 
\begin{figure}
	\centering
		\includegraphics[width=9cm, height=6cm]{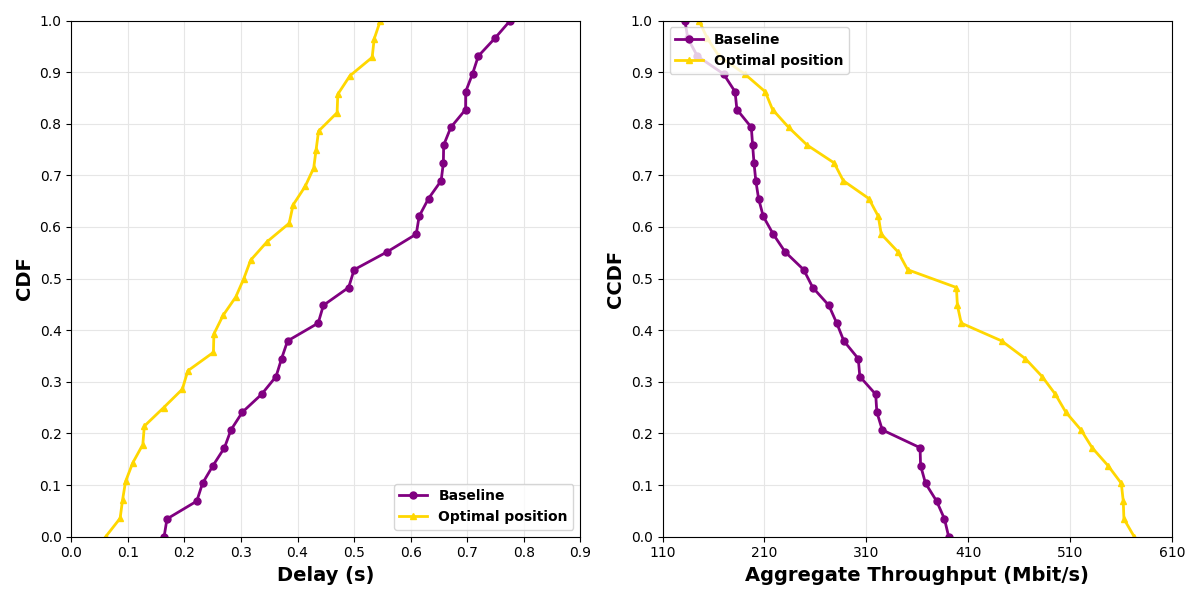}
	\caption{Scenario A: Free-space aggregate throughput and mean delay measured on UAV, where $\lambda_0 =  \lambda_1 =  . . .  =  \lambda_{19}$. .}
   \label{fig3.1}
\end{figure}

\begin{table}[ht]
    \caption{\textbf{Detail of RL parameters.}}
    \label{tab2}
    \setlength{\tabcolsep}{3pt}
    \begin{tabular}{p{100pt} p{130pt}}
        \hline
        Parameter & Amount \\  
        \hline    
        Number of training episodes & 10\\
        $w_1$ & 0.8\\
        $w_2$ & 0.2\\
        Number of evaluation episodes & 1\\        
        Duration of episodes& 100 s\\
        Decision interval, $t_k$ & 100 ms\\       
        Observations & ($x, y, z$), nLoS \\
        Action Space & one-dimensional discrete scaled integer \\
        ML library & TensorFlow\\
        Optimizer & Adam (learning rate of $10^{-2}$) \\
        Epsilon Greedy & 1 (random decision) \\ 
        Quadratic Loss & Mean Square Error (MSE) \\
        Q-function& Two fully connected layers, each with 32 units\\
        Memory Replay & buffer size is $10^6$ with a batch of 64\\
       \hline      
    \end{tabular}
\end{table}

\begin{table}[ht]
    \caption{\textbf{ ns-3 environment configuration }}
    \label{tab1}
    \setlength{\tabcolsep}{3pt}
    \begin{tabular}{p{120pt} p{100pt}}
        \hline
        Parameter & Amount  \\  
        \hline
        Size of venue ($W\times D)$  & 100m  \\        
        Guard Interval $(GI)$ & 800 $ns$\\
        Wi-Fi channel & 50 \\
        Wi-Fi Standard & IEEE 802.11ac  \\
        Channel Bandwidth & 160 MHz  \\
        Antenna Gain & 0dBi  \\
        Tx Power& 20 $dBm$\\
        Noise Floor Power & -85 $dBm$\\
        LoS Propagation Loss Model & ItuR1411LosPropagationLossModel   \\
        NLoS Propagation Loss Model & ItuR1411NlosOverRooftopP ropagationLossModel  \\
        Remote Station Manager mechanism & IdealWifiManager  \\
        Application Traffic & UDP constant bitrate \\        
        Packet Size & 1400 bytes  \\
        \hline
    \end{tabular}
\end{table}   

In Scenario A, the initial position of the UAV \((x_i, y_i, z_i)\), which serves as the baseline, is a 3D coordinate at the center of the venue with an altitude of 10m. In Scenarios B and C, the initial UAV position and baseline are set at 5m above the central building, where a fixed base station is mounted. The configuration of the environment is summarized in Table \ref{tab1}. RLpos-3 users can define these parameters using the \emph{Input/output} module through the $JSON$ file or the command-line interface. A potential zone for UAV placement—referred to as the action space—has been defined to cover all areas above the venue. This zone can be configured in the $JSON$ file based on the requirements of the specific scenario. Furthermore, the observation space is defined by the UAV's position \((x, y, z)\) and the number of LoS connections between the UAV and UEs \(nLoS\) at each time step. The reward function is formulated in Equation \ref{reward} to maximize $nLoS$ and throughput with different weights. Additionally, RLpos-3 allows users to customize the reward function by setting up the \emph{gym} module. The main configuration parameters are detailed in Table \ref{tab2}.

\begin{equation}\label{reward}
    \begin{aligned}
        & r = w_1 \cdot nLoS_{\text{norm}} + w_2 \cdot Throughput_{\text{norm}} & \\
        & \text{where} \quad nLoS_{\text{norm}} = \frac{nLoS}{N} & \\
        &  \quad\quad\quad Throughput_{\text{norm}} = \frac{Throughput}{\sum_{i=0}^{N} \lambda_i} &
    \end{aligned}
\end{equation}

where $N$ is the number of UEs.

\subsection{Simulation Results~\label{sec:simulation_Results}}

The optimal position achieved by RLpos-3 is evaluated in this section using ns-3. The results are derived from 30 simulations, while the algorithm is trained over 10 episodes, each lasting 100 seconds. All simulations were conducted using the \(SetRandomSeed()\) function and \(RngRun = \{1, 2, \dots, 30\}\) parameters under constant networking conditions. The aggregate throughput achieved by the UAV is illustrated using the complementary cumulative distribution function (CCDF) and the mean delay is depicted using the cumulative distribution function (CDF).

\begin{figure}
	\centering
           \includegraphics[width=8cm, height=6cm]{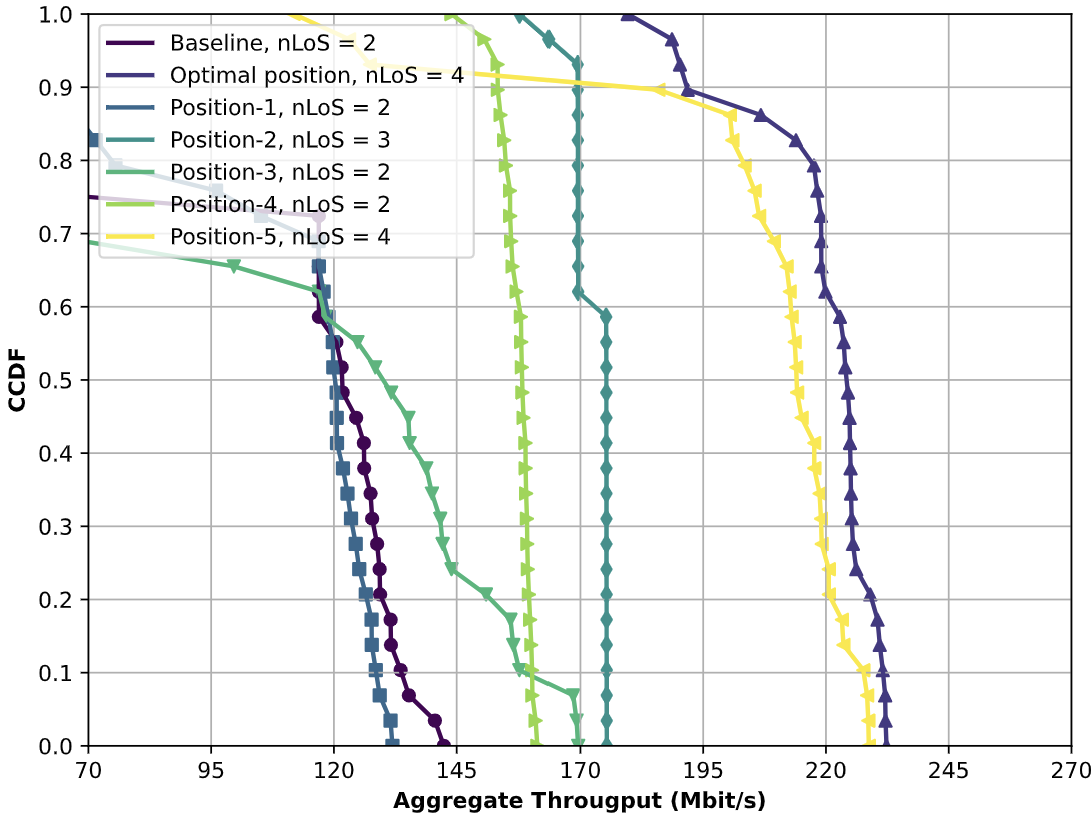}
	\caption{Scenario B: Aggregate throughput measured on UAV, where $\lambda_0 =  \lambda_1 =  \lambda_2 =  \lambda_3$.}
	\label{fig4}
\end{figure}
\begin{figure}
	\centering
		\includegraphics[width=8cm, height=6cm]{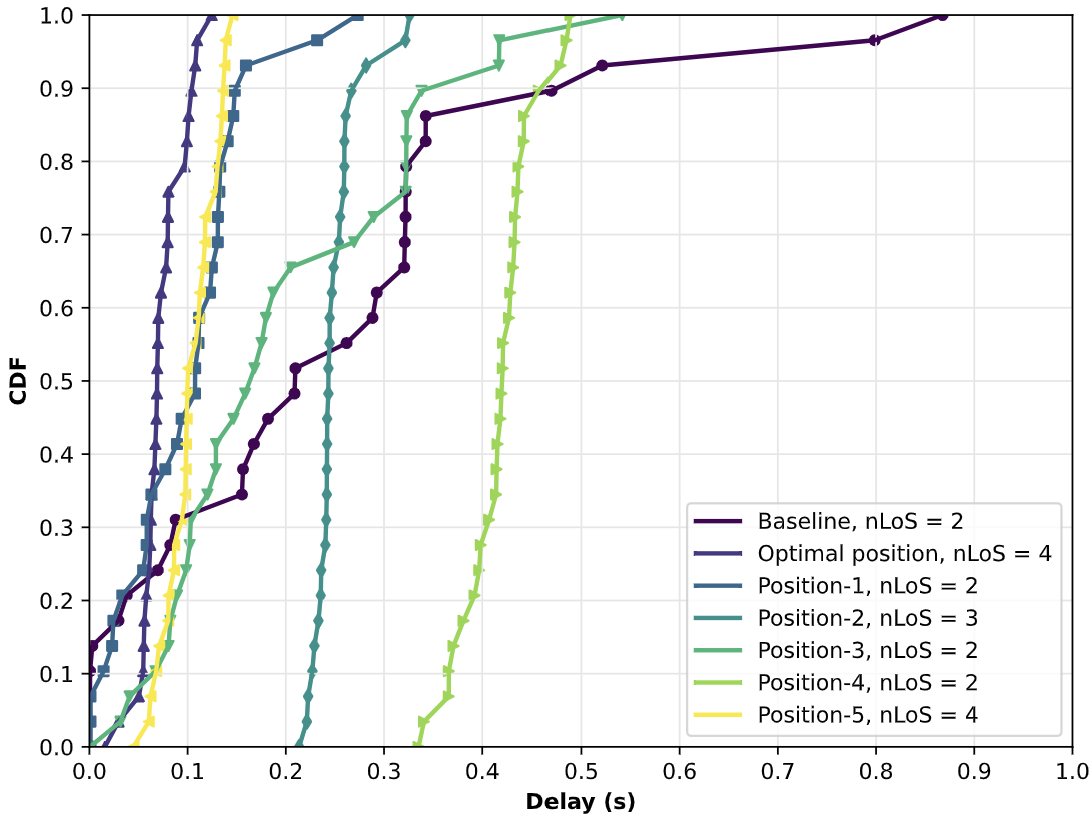}
	\caption{Mean delay measured on UAV, where $\lambda_0 = \lambda_1 = \lambda_2 = \lambda_3$.}
	\label{fig5}
\end{figure}

\begin{figure}
	\centering
		\includegraphics[width=8cm, height=6cm]{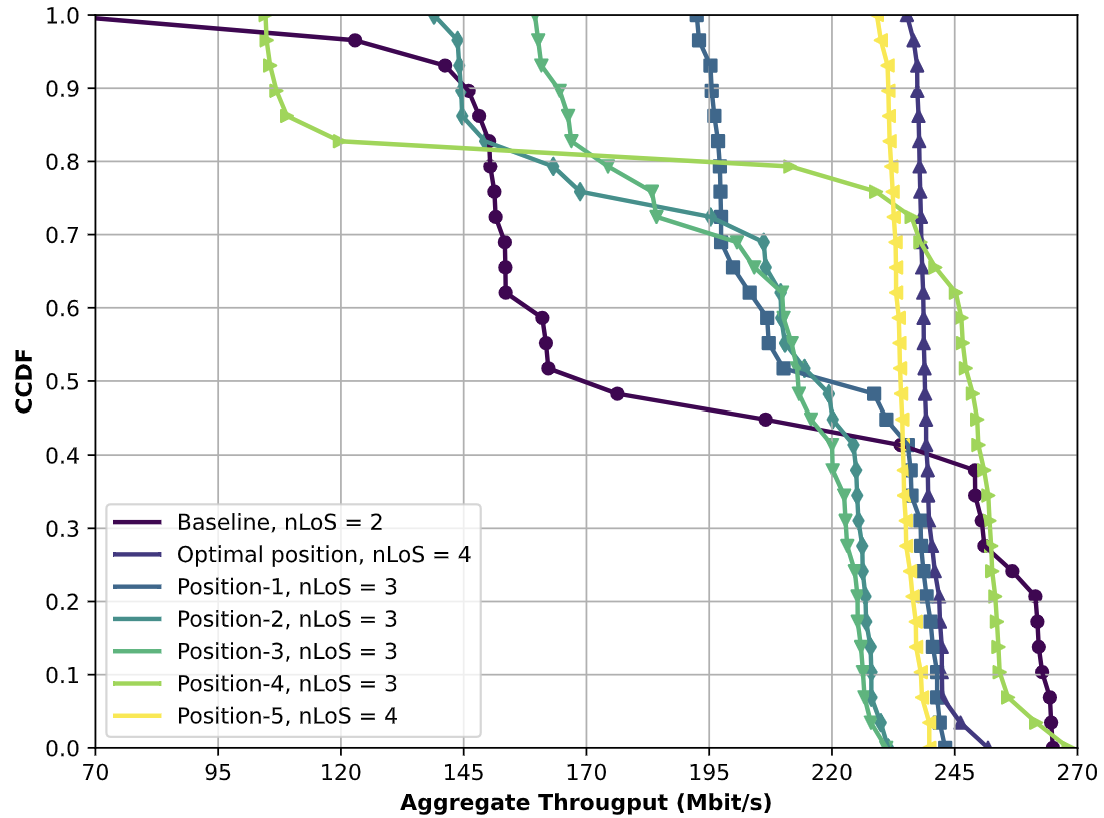}
	\caption{Aggregate throughput measured on UAV, where $\lambda_0 = 0.75 \times \lambda_1 = 2 \times \lambda_2 = 4 \times \lambda_3$.}
	\label{fig6}
\end{figure}

\begin{figure}
	\centering
		\includegraphics[width=8cm, height=6cm]{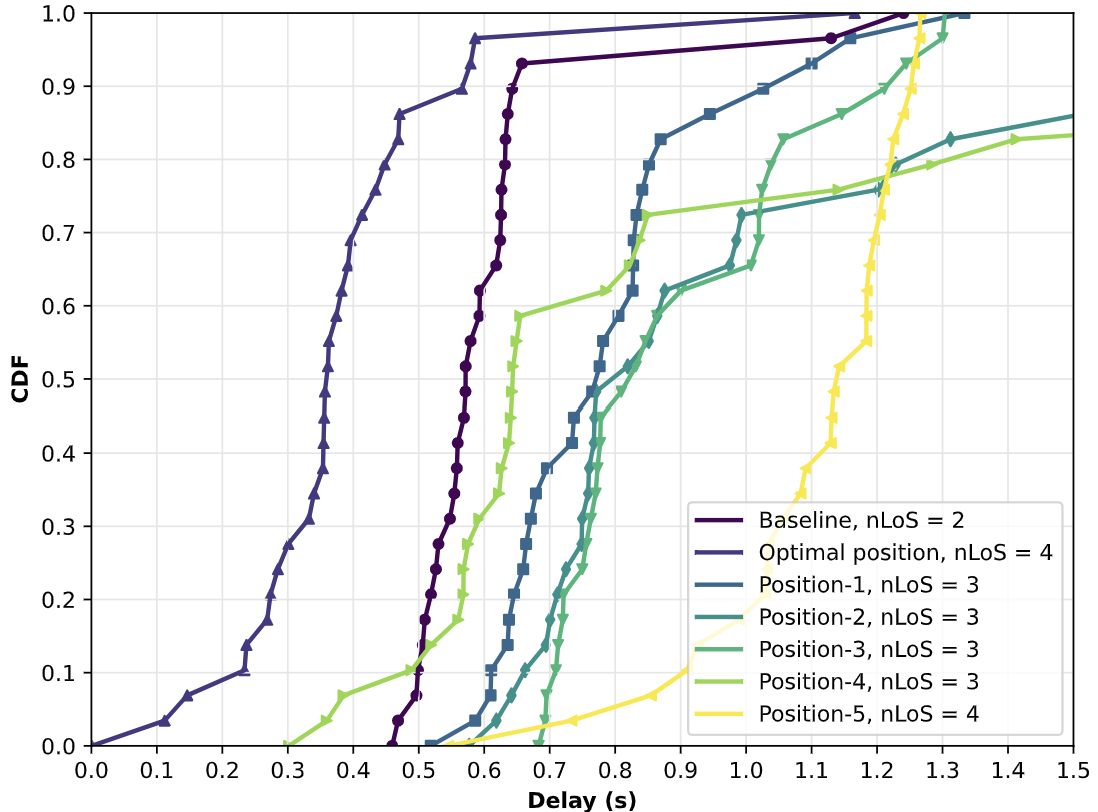}
	\caption{Mean delay measured on UAV, where $\lambda_0 = 0.75 \times \lambda_1 = 2 \times \lambda_2 = 4 \times \lambda_3$.}
	\label{fig7}
\end{figure}

Figure \ref{fig3.1} shows the evaluation results for Scenario A, where the aggregate throughput and delay at the optimal position achieved by RLpos-3 are compared with the baseline. The results indicate a 60\% improvement in throughput and a 40\% reduction in mean delay. For both Scenarios A and B, five additional positions—each located 10 meters away from the optimal position in different directions, denoted as positions 1–5—are evaluated against the optimal position determined by RLpos-3 and the baseline. Figures \cref{fig4} and \cref{fig5} illustrate the results of Scenario B, showing an 80\% increase in throughput and a 60\% reduction in delay. In heterogeneous Scenario C, the framework achieved a 60\% improvement in aggregate throughput and a 35\% reduction in median delay compared to other positioning methods as depicted in Figures \ref{fig6} and \ref{fig7}.

It is worth noting that, due to the multi-objective reward function, in Scenarios B and C, even though Position 4 maintains LoS with all UEs, the optimal position identified by RLpos-3 still achieves higher throughput and lower delay.

\section{Conclusions~\label{sec:Conclusions}}

RLpos-3 is a simulation framework designed to implement, validate, and evaluate adaptive, obstacle-aware, RL-based UAV positioning algorithms. It integrates existing reinforcement learning libraries with ns-3 by leveraging ns-3gym as a bridging interface. RLpos-3 supports standard RL libraries such as TensorFlow Agents and OpenAI Gym, customizing them to align with the specific requirements of UAV positioning algorithms across various target scenarios. By simply configuring the \emph{Input/Output} module, users can develop and evaluate positioning strategies tailored to diverse environments. The framework aims to optimize UAV placement to enhance network performance—most notably by increasing aggregate throughput and reducing delay. \looseness=-1
The results from a simple use case illustrate RLpos-3's potential as a valuable tool for advancing RL-based UAV positioning solutions. Future work will focus on integrating signal processing techniques and computer vision methods to detect the positions of ground users and obstacles. Furthermore, the framework will be extended to support multi-UAV scenarios.


\bibliographystyle{IEEEtran}
\bibliography{IEEEabrv,References}

\begin{thebibliography}{10}
\providecommand{\url}[1]{#1}
\csname url@samestyle\endcsname
\providecommand{\newblock}{\relax}
\providecommand{\bibinfo}[2]{#2}
\providecommand{\BIBentrySTDinterwordspacing}{\spaceskip=0pt\relax}
\providecommand{\BIBentryALTinterwordstretchfactor}{4}
\providecommand{\BIBentryALTinterwordspacing}{\spaceskip=\fontdimen2\font plus
\BIBentryALTinterwordstretchfactor\fontdimen3\font minus \fontdimen4\font\relax}
\providecommand{\BIBforeignlanguage}[2]{{%
\expandafter\ifx\csname l@#1\endcsname\relax
\typeout{** WARNING: IEEEtran.bst: No hyphenation pattern has been}%
\typeout{** loaded for the language `#1'. Using the pattern for}%
\typeout{** the default language instead.}%
\else
\language=\csname l@#1\endcsname
\fi
#2}}
\providecommand{\BIBdecl}{\relax}
\BIBdecl

\bibitem{shafafi2023uav}
K.~Shafafi, E.~N. Almeida, A.~Coelho, H.~Fontes, M.~Ricardo, and R.~Campos, ``Uav-assisted wireless communications: An experimental analysis of a2g and g2a channels,'' in \emph{Simulation Tools and Techniques}.\hskip 1em plus 0.5em minus 0.4em\relax Cham: Springer Nature Switzerland, 2024, pp. 243--256.

\bibitem{shafafi2024traffic}
K.~Shafafi, M.~Ricardo, and R.~Campos, ``Traffic and obstacle-aware uav positioning in urban environments using reinforcement learning,'' \emph{IEEE Access}, vol.~12, pp. 188\,652--188\,663, 2024.

\bibitem{shafafi2023joint}
K.~Shafafi, A.~Coelho, R.~Campos, and M.~Ricardo, ``Joint traffic and obstacle-aware uav positioning algorithm for aerial networks,'' in \emph{2023 IEEE 9th World Forum on Internet of Things (WF-IoT)}, 2023, pp. 1--6.

\bibitem{jx7c-py87-25}
\BIBentryALTinterwordspacing
K.~Shafafi, ``Uav positioning and optimization framework with obstacle-aware modeling using gekko,'' 2025. [Online]. Available: \url{https://dx.doi.org/10.21227/jx7c-py87}
\BIBentrySTDinterwordspacing

\bibitem{rs11121443}
\BIBentryALTinterwordspacing
H.~Yao, R.~Qin, and X.~Chen, ``Unmanned aerial vehicle for remote sensing applications—a review,'' \emph{Remote Sensing}, vol.~11, no.~12, 2019. [Online]. Available: \url{https://www.mdpi.com/2072-4292/11/12/1443}
\BIBentrySTDinterwordspacing

\bibitem{8875210}
L.~Shi, N.~J.~H. Marcano, and R.~H. Jacobsen, ``A survey on multi-unmanned aerial vehicle communications for autonomous inspections,'' in \emph{2019 22nd Euromicro Conference on Digital System Design (DSD)}, 2019, pp. 580--587.

\bibitem{Reinforc82:online}
\BIBentryALTinterwordspacing
``Reinforcement learning (dqn) tutorial — pytorch tutorials 2.7.0+cu126 documentation,'' [Online; accessed 2025-06-11]. [Online]. Available: \url{https://docs.pytorch.org/tutorials/intermediate/reinforcement_q_learning.html}
\BIBentrySTDinterwordspacing

\bibitem{TensorFl16:online}
\BIBentryALTinterwordspacing
``Tensorflow,'' [Online; accessed 2025-02-12]. [Online]. Available: \url{https://www.tensorflow.org/}
\BIBentrySTDinterwordspacing

\bibitem{GitHubop15:online}
\BIBentryALTinterwordspacing
openai, ``Github - openai/gym: A toolkit for developing and comparing reinforcement learning algorithms.'' [Online; accessed 2025-01-23]. [Online]. Available: \url{https://github.com/openai/gym}
\BIBentrySTDinterwordspacing

\bibitem{ns3adisc37:online}
``ns-3 | a discrete-event network simulator for internet systems,'' \url{https://www.nsnam.org/}, (Accessed on 01/09/2024).

\bibitem{GitHubtk28:online}
``Github - tkn-tub/ns3-gym: ns3-gym - the playground for reinforcement learning in networking research,'' \url{https://github.com/tkn-tub/ns3-gym}, (Accessed on 01/09/2024).

\bibitem{RLpos3Re43:online}
\BIBentryALTinterwordspacing
K.~Shafafi, ``Rlpos-3: Reinforcement learning for positioning,'' 2025. [Online]. Available: \url{https://dx.doi.org/10.21227/rp8v-jd54}
\BIBentrySTDinterwordspacing

\bibitem{8761897}
J.~He, J.~Wang, H.~Zhu, W.~Cheng, P.~Yue, and X.~Yi, ``Resource allocation in drone aided emergency communications,'' in \emph{ICC 2019 - 2019 IEEE International Conference on Communications (ICC)}, 2019, pp. 1--6.

\bibitem{8422376}
B.~Galkin, J.~Kibilda, and L.~A. DaSilva, ``Backhaul for low-altitude uavs in urban environments,'' in \emph{2018 IEEE International Conference on Communications (ICC)}, 2018, pp. 1--6.

\bibitem{9684492}
T.~Liang, T.~Zhang, J.~Yang, D.~Feng, and Q.~Zhang, ``Uav-aided positioning systems for ground devices: Fundamental limits and algorithms,'' \emph{IEEE Internet of Things Journal}, vol.~9, no.~15, pp. 13\,470--13\,485, 2022.

\bibitem{8369021}
H.~Wang, H.~Zhao, L.~Zhou, D.~Ma, and J.~Wei, ``Deployment algorithm for minimum unmanned aerial vehicles towards optimal coverage and interconnections,'' in \emph{2018 IEEE Wireless Communications and Networking Conference Workshops (WCNCW)}, 2018, pp. 72--277.

\bibitem{7921981}
H.~Shakhatreh, A.~Khreishah, A.~Alsarhan, I.~Khalil, A.~Sawalmeh, and N.~S. Othman, ``Efficient 3d placement of a uav using particle swarm optimization,'' in \emph{2017 8th International Conference on Information and Communication Systems (ICICS)}, 2017, pp. 258--263.

\bibitem{7510820}
R.~I. Bor-Yaliniz, A.~El-Keyi, and H.~Yanikomeroglu, ``Efficient 3-d placement of an aerial base station in next generation cellular networks,'' in \emph{2016 IEEE International Conference on Communications (ICC)}, 2016, pp. 1--5.

\bibitem{ML8121867}
M.~Wang, Y.~Cui, X.~Wang, S.~Xiao, and J.~Jiang, ``Machine learning for networking: Workflow, advances and opportunities,'' \emph{IEEE Network}, vol.~32, no.~2, pp. 92--99, 2018.

\bibitem{MLJiang2017MachineLP}
\BIBentryALTinterwordspacing
C.~Jiang, H.~Zhang, Y.~Ren, Z.~Han, K.-C. Chen, and L.~Hanzo, ``Machine learning paradigms for next-generation wireless networks,'' \emph{IEEE Wireless Communications}, vol.~24, pp. 98--105, 2017. [Online]. Available: \url{https://api.semanticscholar.org/CorpusID:31583964}
\BIBentrySTDinterwordspacing

\bibitem{43-bab6dcd317da4e3d82375669cbb45023}
Y.~Munaye, H.~Lin, A.~Adege, and G.~Tarekegn, ``\BIBforeignlanguage{???core.languages.en_GB???}{Uav positioning for throughput maximization using deep learning approaches},'' \emph{\BIBforeignlanguage{???core.languages.en_GB???}{Sensors}}, vol.~19, no.~12, Jun. 2019, publisher Copyright: {\textcopyright} 2019 by the authors. Licensee MDPI, Basel, Switzerland.

\bibitem{23-8644345}
X.~Liu, Y.~Liu, and Y.~Chen, ``Deployment and movement for multiple aerial base stations by reinforcement learning,'' in \emph{2018 IEEE Globecom Workshops (GC Wkshps)}, 2018, pp. 1--6.

\bibitem{22-COLONNESE2019101872}
S.~Colonnese, F.~Cuomo, G.~Pagliari, and L.~Chiaraviglio, ``Q-square: A q-learning approach to provide a qoe aware uav flight path in cellular networks,'' \emph{Ad Hoc Networks}, vol.~91, p. 101872, 2019.

\bibitem{MCSTable16:online}
``Mcs table and how to use it – wireless lan professionals,'' \url{https://wlanprofessionals.com/mcs-table-and-how-to-use-it/}, (Accessed on 01/15/2025).

\end{thebibliography}

\end{document}